\documentclass[letterpaper]{article} 
\usepackage[submission]{aaai23}  
\usepackage{times}  
\usepackage{helvet}  
\usepackage{courier}  
\usepackage[hyphens]{url}  
\usepackage{graphicx} 
\usepackage{natbib}  
\usepackage{caption} 
\usepackage{algorithm}
\usepackage{algorithmic}

\usepackage{newfloat}
\usepackage{listings}
\DeclareCaptionStyle{ruled}{labelfont=normalfont,labelsep=colon,strut=off} 
\floatstyle{ruled}
\newfloat{listing}{tb}{lst}{}
\floatname{listing}{Listing}
\title{EasyRec: An easy-to-use, extendable and efficient framework for building industrial recommendation systems}
\author {
    Mengli Cheng,
    Yue Gao,
    Guoqiang Liu,
    HongSheng Jin,
    Xiaowen Zhang
}
\affiliations {
    Alibaba Group \\
    \{mengli.cml, dawn.gy, yancheng.lgq, hongsheng.jhs, zxw320697\}@alibaba-inc.com
}
\begin{document}

\maketitle

\begin{abstract}
We present EasyRec, an easy-to-use, extendable and efficient recommendation framework for building industrial recommendation systems.
Our EasyRec framework is superior in the following aspects:
first, EasyRec adopts a modular and pluggable design pattern to reduce the efforts to build custom models;
second, EasyRec implements hyper-parameter optimization and feature selection algorithms to improve model performance automatically;
third, EasyRec applies online learning to fast adapt to the ever-changing data distribution.
The code is released: https://github.com/alibaba/EasyRec.
\end{abstract}

\maketitle

\section{Introduction}
Recommendation systems are widely used in customer-oriented business platforms as they improve user engagement and platform revenues.
%
However, building practical recommendation systems requires large amounts of effort.
Due to the large number of items, recommendations system usually adopts "funnel approach"~\cite{Covington2016DeepNN}, which consists of multiple stages: candidate matching, pre-ranking, ranking, and probably re-ranking.
Each of the stages requires building one or several models.
What's more, building the models usually requires training, evaluation, export, and online serving, as illustrated in Fig.\ref{fig:framework}(b).
Meanwhile, deep models are computation intensive and require large amounts of training data, therefore, distributed training is necessary for fast convergence.
The huge amounts of computation also brings challenges for deploying efficient online serving system.
Besides, offline and online features often suffer from inconsistencies, as their generation process may be maintained by different people.
In this demo, we present EasyRec, an easy-to-use recommendation framework to address the above challenges.
%


\section{Overview}

\begin{figure}
        \centering
        \includegraphics[width=\columnwidth]{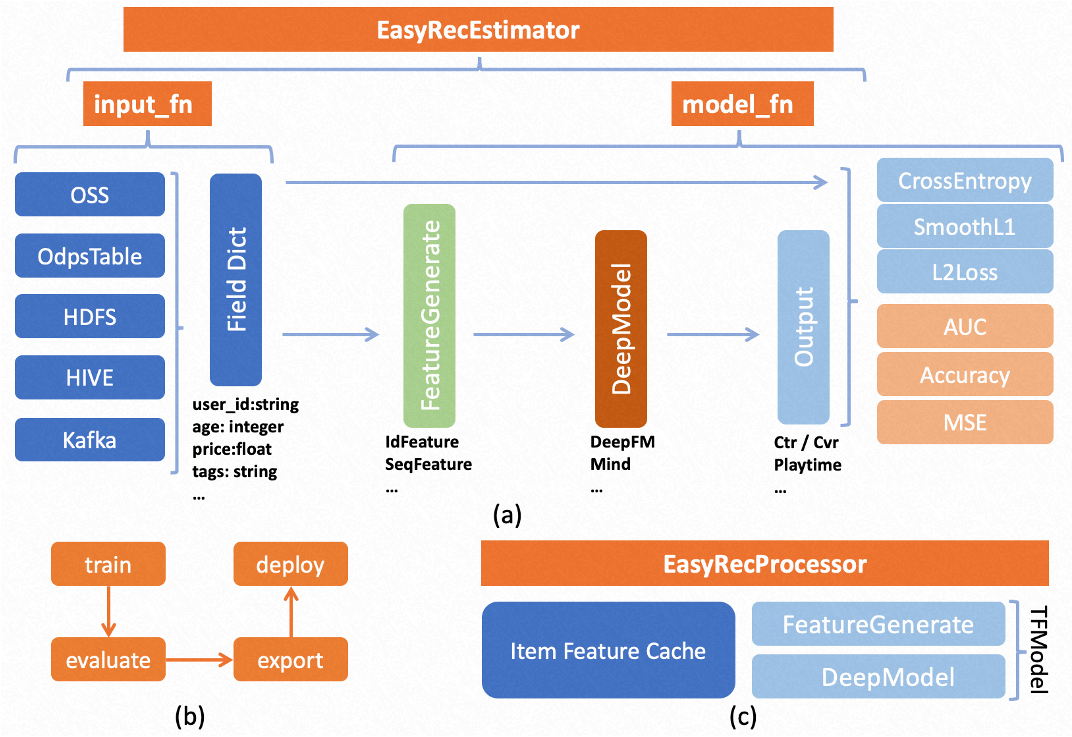}
        \caption{An overview framework of EasyRec: (a) architecture, (b) model development workflow, (c) online serving. }
        \label{fig:framework}
\vspace{-2mm}
\end{figure}

In this section, we introduce the framework of EasyRec in detail.
As illustrated in Fig.\ref{fig:framework}(a), EasyRec adopts a modular design, and the modules work in a pluggable mode.
Thanks to the pluggable design, each module could be extended separately and flexibly without modifying any other modules.
For example, to build a new Click-Through-Rate(CTR) estimation model, we only need to define the network structure and related parameters.
Then the new CTR model can run on all distributed platforms supported by EasyRec, such as the Machine Learning Platform for AI~(PAI) of Alibaba Cloud~\footnote{https://www.alibabacloud.com/en/product/machine-learning}.
EasyRec consists of the following modules: data loading module, feature generation module, deep learning model module, loss and metric function module.
All the modules are configurable with hyper-parameters defined in protobuf text format.
In data module, data readers are implemented to support most big data engines, including hive, oss, kafka, etc. 
EasyRec also supports multiple data formats such as csv, tfrecord, parquet, etc. 
The core of feature generation module is implemented using C++ for high performance, which is then wrapped with tensorflow operators so that it could be easily integrated into tensorflow graphs.
The feature operators are re-used in online serving to ensure consistency between online and offline features.
Moreover, EasyRec integrates more than 20 state-of-the-art models for different stages of recommendation systems, such as deep structured semantic models~\cite{Huang2013LearningDS} for candidate matching and deep interest network~\cite{Zhou2018DeepIN} for ranking.
We further adapt the tensorflow estimator~\cite{Cheng2017TensorFlowEM} to support distributed training and evaluation on multiple platforms(Kubernetes~\cite{Kubeflow}, Yarn~\cite{Yarn}, etc), which is named EasyRecEstimator. 

\section{Model Train}
EasyRec provides easy-to-use interfaces for multiple distributed platforms.
For example, we could submit a distributed training job on PAI with just a simple command:
\begin{quote}
{\small
\begin{verbatim}
pai -name easy_rec_ext -Dcmd=train
-Dconfig=oss://easyrec/test/deepfm.config
-Dtrain_tables=odps://test/tables/train
-Deval_tables=odps://test/tables/test
-Dcluster='{"ps":{"count":1,"cpu":1000}, 
  "worker":{"count":3,"cpu":800}}'
-Dmodel_dir=oss://easyrec/models/deepfm/
\end{verbatim}
}
\end{quote}

Here, we created a distributed training job of DeepFM~\cite{Guo2017DeepFMAF} with 1 parameter server and 3 workers.
%
The configuration file~\texttt{deepfm.config} defines all the parameters,
which consists of~\texttt{data\_config}, \texttt{feature\_config}, \texttt{model\_config}, \texttt{train\_config}, and~\texttt{eval\_config}.

EasyRec also provides graphical interfaces for the Kubernetes platform to start training jobs with just a few clicks~\footnote{easyrec.readthedocs.io/en/latest/quick\_start/dlc\_tutorial.html}.
%
EasyRec supports two kinds of distribution strategies: parameter server strategy and multi-worker mirrored strategy.
We further optimize the later with sparse operation kit~\cite{SOK}, which speeds up training process by more than 3 times on criteo dataset~\footnote{www.kaggle.com/competitions/criteo-display-ad-challenge}.

\section {Hyper-Parameter Optimization}
Recommendation models are usually faced with data-sparsity problem~\cite{Li2019MultiInterestNW} due to the high-dimensional feature space of categorical features.
Therefore, the hyper-parameters must be carefully tuned to avoid over-fitting over training data.
In EasyRec, we address this problem by providing hyper-parameter optimization through seamless integration with NNI~\cite{NNI}.
We also employ early stopping strategies~\cite{Golovin2017GoogleVA} to improve search efficiency.
%
As described above, EasyRec defines all the hyper-parameters in protobuf text format and optimizes them with simple definitions of parameter names and search spaces.
%
%
For example, we could optimize the regularization parameter on feature embedding with the following config:

\begin{quote}
{\small
\begin{verbatim}
"model_config.embedding_regularization":
{"_type":"uniform","_value":[1e-6, 1e-4]}
\end{verbatim}
}
\end{quote}

\section {Feature Selection}
Typical industrial recommendation systems usually generate hundreds to thousands of features.
Too many features not only cause the data-sparsity problem, but also require large amounts of computation. 
EasyRec provides systematic methods to evaluate the importance of features, and filter out features of lower importance.
The importance of features is evaluated by variational dropout~\cite{Ma2021TowardsAB}, which is improved by optimizing the dropout hyper-parameters for both training and test data, preventing over-fitting over training data, and leading to better performance on test data.
%
%
We apply our improved method to a typical scene with more than 1000 features: about half of the features are deleted, and the inference speed of online serving is improved by 2 times without degrading performance.

\section {Online Serving}
Online serving faces two main challenges: i)~large numbers of Queries Per Second(QPS); ii)~Request Time(RT) is usually limited to less than 0.5 seconds.
EasyRec optimizes online serving performance by the following techniques:
%
first, a LRU cache is maintained to ensure fast access to frequent items; 
second, we integrate feature generation into tensorflow graph by wrapping them as tensorflow operators, leading to interleaved execution among different features;
%
%
third, we optimize embedding lookup with vector instructions to speed up embedding addition and multiplication operations.
With the above optimizations, the RT of online serving is reduced to 1/4 of the original implementations.

\section {Online Learning}
Online learning improves performance by fast adaption to the changing distribution of features and targets, which is caused by the introduction of new items and users.
However, deploying an online learning service mainly faces two challenges:
first, a real-time sample streams must be created;
second, deep models are generally too large, making fast updating of whole models difficult.
EasyRec solves the first challenge by building a flink-based template to create real-time sample streams, which consists of three modules: events aggregation, label generation, join of label and feature stream.
EasyRec overcomes the second challenge by incremental update.
Specifically, we record the parameters updated during training, and periodically send changed parameters to message queues.
Online service retrieves changes from the message queues, and applies them to EasyRec models.
In a typical scene, the number of parameters updated is reduced to less than 5\% of all parameters.
The incremental updates are applied within seconds, compared to the 10+ minutes delay in updating all parameters.

\section {Related Works}
Previously, many recommendation model libraries are released, such as FuxiCtr~\cite{Zhu2020BarsCTROB}, TorchRec~\cite{TorchRec}, and DeepCtr~\cite{shen2017deepctr}.
EasyRec differs from them in that it is not just a model library, it also provides deeply optimized and easy-to-use training and serving services on distributed platforms. It further applies hyper-parameter optimization, feature selection, and online learning to generate high performance models with efficiency.

\section {Conclusion}
In this demo, we present EasyRec, an easy-to-use, extendable, and efficient recommendation framework which runs on many distributed platforms.
It provides hyper-parameter tuning and feature selection for automatic model optimization.
It further implements online learning to fast adapt to the ever-changing data distribution.
In the future, we will continue to develop our framework to support more state-of-the-art recommendation models.
\newpage

\bibliography{easy_rec.bib}


\end{document}